\documentclass[]{spie}  

\usepackage{amsmath,amsfonts,amssymb}
\usepackage{graphicx}
\usepackage[colorlinks=true, allcolors=blue]{hyperref}

\title{Development and potential of new Micro-Mirror Devices optimized for astronomy}

\author[a,b,*]{Robberto, M.}
\author[c]{Gong. C.}
\author[c]{Huffman, J.}
\author[d]{Ninkov, Z.}
\author[d]{Puchades, I.}
\author[a,b]{Gennaro, M.}
\author[a,b]{Kassin, S. A.}
\author[b]{Smee, S. A.}
\affil[a]{Space Telescope Science Institute, 3700 San Martin Dr., Baltimore, MD 21218, USA}
\affil[b]{Johns Hopkins University, Bloomberg Center for Physics and Astronomy, 3400 N. Charles Street, Baltimore, MD 21218, USA}
\affil[c]{Texas Christian University, Department of Engineering, 2840 West Bowie Street, Fort Worth, Texas 76109, USA}
\affil[d]{Rochester Institute of Technology, Center for Imaging Science, 54 Lomb Memorial Drive, Rochester, NY 14623-560, USA}

\authorinfo{Further author information: (Send correspondence to M.R., C. Gong)\\M.R.: E-mail: robberto@stsci.edu, Telephone: 1 (410) 338-4382; C. G.: E-mail: c.gong@tcu.edu, Telephone: +1 (817)257-6317}

\begin{document} 
\maketitle
\begin{abstract}
We introduce our new program to develop two-dimensional MEMS arrays of individually addressable micro-mirrors (``Micro-Mirror Devices'', MMDs) specifically optimized for astronomy, multi-slit spectroscopy in particular. After reviewing the main characteristics and performance of the currently available options, Micro Shutter Arrays by NASA/Goddard and Digital Micromirror Devices by Texas Instruments, we present our planned first generation/baseline devices with 30\textmu m x 30\textmu m pixel size arranged in a 1K x 1K format with tilt angle 15 degrees. Our goal is to bring to maturity a technology capable of delivering arrays of 2K x 2K element of 100\textmu m x 100\textmu m, buttable on two sides to achieve even larger formats. In additions to MEMS design, we will develop the associated device packaging and electronic control circuitry leveraging on the extensive expertise gained in the last 30+ years by leading experts from digital imaging industry. 
\end{abstract}


\section{INTRODUCTION}
\label{sec:intro}  
Surveys of the sky across multiple spectral bands from ground- and space-based observatories are central to progress in astronomy.  Multi-object spectrographs (MOS), in particular, allows astronomers to acquire simultaneously the spectra of many target objects spread across a field of view. MOS provide enormous efficiency gains since astronomical measurements normally require long integration time to achieve a desired signal-to-noise ratio (SNR). 
Commonly used multiplexing mechanisms include pre-fabricated slit masks, reconfigurable slit units, and optical fiber positioners, each approach presenting peculiar advantages and disadvantages. Pre-fabricated slit masks have to be machined weeks in advance into a plate based on the preliminary imaging of the object field to be observed. Observing conditions and long-term opto-mechanical stability must be taken into account, and each pointing requires exchanging and realigning a different mask.  Instrument like Gemini Multi-Object Spectrographs (GMOS) allow up to 18 masks per night with typically 50 targets (galaxy) each.  Reconfigurable slit units consist of sliding bars that work in pairs to create slits. While sliding bars units provide a significant gain in efficiency due to a faster reconfiguration time, they usually have a limit number of slits, e.g. 46 slits (i.e. targets) for Keck/MOSFIRE.  Fiber-based MOS uses optical fibers that are precisely placed at the focal plane of a telescope in locations corresponding to target objects. This solution has been adopted by mid-class and large telescopes like SDSS, Apogee, SUBARU/PFS. The latest robotic fiber positioning systems have greatly improved reconfiguration times, but  these systems are large, complex, and expensive. 

\begin{figure}[tph!]
\centerline{\includegraphics[trim={4.3cm 19.5cm 5cm 3.5cm}, totalheight=4.5cm]{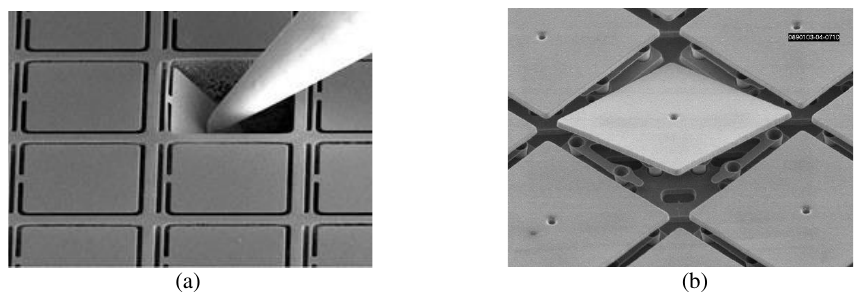}}
    \caption{(a) Microshutter Array (NASA website); (b) Digital micro-mirror by Texas Instruments.}
    \label{fig:MSA DMD}
\end{figure}

In the last two decades, a different approach based on micro-electro-mechanical systems (MEMS) devices has been considered. MEMS can be designed as rapidly re-configurable slit masks for MOS. There are two different approaches:  Microshutter arrays (MSAs) and digital micromirror devices (DMDs) as shown in Figure \ref{fig:MSA DMD}.  Both MEMS devices are comprised of an array of individually addressable elements that act as optical switches to selectively capture the spectra of target sources. MSAs have transmissive elements (microshutters) that are opened or closed to selectively transmit light to a spectrograph. DMDs instead have reflective elements (micromirrors) that direct light either towards or away from the spectrograph. While MSA have been created and adopted only by NASA for the NIRSpec spectrograph onboard the James Webb Space Telescope, there are several DMD MOS that have been deployed on ground-based telescopes including RITMOS\cite{meyer_ritmos_2004}, IRMOS\cite{mackenty_irmos_2003} and SAMOS\cite{evans_samos_2016}. MEMS-based MOS offer significant advantages over other more conventional multiplexing approaches, the main ones being the possibility of rapid and accurate positioning of the slit masks, high spatial density, repeatability, compact size, and modularity, intended as the ability to group adjacent elements to adjust slit size to vary spectral resolution and maximize signal-to-noise in the spectral direction while capturing local sky  variations in the cross-dispersed direction. Moreover, MEMS-based MOS can perform IFU-like observations on multiple targets through slitlet stepping\cite{kassin_s_slitlet_2025} or Hadamard Transform spectroscopy\cite{FixsenTransform, oram_modeling_2022}.

In this paper we present our MMD development plan, led by former TI engineers who worked on the development of the DMDs, aimed at developing a micro mirror device that will maintain the extraordinary reliability of TI devices but optimized for astronomical applications.  Our new micro mirror devices will have a larger mirror size of 30\textmu m (increase from the TI largest mirror of 13.7\textmu m), higher tilt angle 15\textdegree (increase from 12\textdegree),  more mirrors (with a goal of 4 million) to deliver a larger field of view (FOV) (e.g. $\sim 13\times 13$ arcmin vs. current $3\times3$ arcmin assuming the same design parameters of SAMOS at the SOAR telescope). It will also have superior throughput as astronomers will be able to optimize mirror coating and window materials to the wavelength range of interest. In general, our technology will allow producing devices optimized for specific applications, from wide-field surveys to extreme adaptive optics.  The new device will drastically lower the cost of  the MOS system by employing much simpler mirror operation control circuitry and MEMS fabrication and packaging processes. In summary our goal is to develop a robust, low-cost, functional micro mirror device that will enable a new generation of  MOS with unprecedented capabilities, a major innovation that will impact nearly every major field of astrophysics.

\section{Assessment of existing MEMS technology for MOS spectroscopy}

The two leading technologies, MSA and DMD, have specific advantages and limitations. For example, only MSAs have 100\% transmission efficiency, but only DMDs allow to acquire simultaneously an image of the same field targeted by the spectrograph, thus maximizing observing efficiency. In what follows we summarize the main characteristics of the two types of devices in order to clarify the key factors driving the requirement on our new family of MMD arrays. 


\subsection{MSAs}
The development of MSAs by a team at the NASA/GSFC has been driven by the original JWST mission requirements, including contrast $> 2000:1$, lifetime $>9.4\times 10^4$ cycles and operating
temperature down to 35~K\cite{moseley_microshutters_2004}. A JWST/NIRSPEC MSA 
is a grid of $365\times 171$ rectangular shutters measuring $105~\mu$m$\times 204~\mu$m center-to-center. The shutters blades are silicon nitride membranes supported by a torsion bar hinge on the long side.  The individual shutters are magnetically activated and electrostatically latched. The blades are lined with magnetic strips. In their resting state, the torsion bar keeps the blade in the closed position. To activate the shutter, an initial magnet sweep repels the magnetic strips tilting all blades in the open position.  The walls of the individual metal boxes can be electrically charged to hold the blade open. Reversing the wall voltage allows the blades to return to the original position, after a second pass of the magnet. By individually addressing each cell, a binary mask of open and closed shutters can be synthesized\cite{li_mems_2008}.

Building on the experience gained developing the original MSAs, around 2010 NASA/GSFC started the development of the Next Generation Microshutter Arrays (NGMSA). NGMSA avoid the magnetic activation steps and are operated exclusively via electrostatic forces. In addition, NGMSA implement a number of modifications intended to improve device reliability \cite{li_jwst_2010,li_successful_2020}. A research group at the University of Tokyo has also investigated the development of MSAs for ground-based MOS instruments adopting an approach similar to the one pioneered by the NASA/GSFC for JWST\cite{liu_random_2020}. 

\subsection{DMDs}
\begin{figure}[tph!]
\centerline{\includegraphics[trim={2cm 11.5cm 3cm 2.5cm}, totalheight=7.5cm]{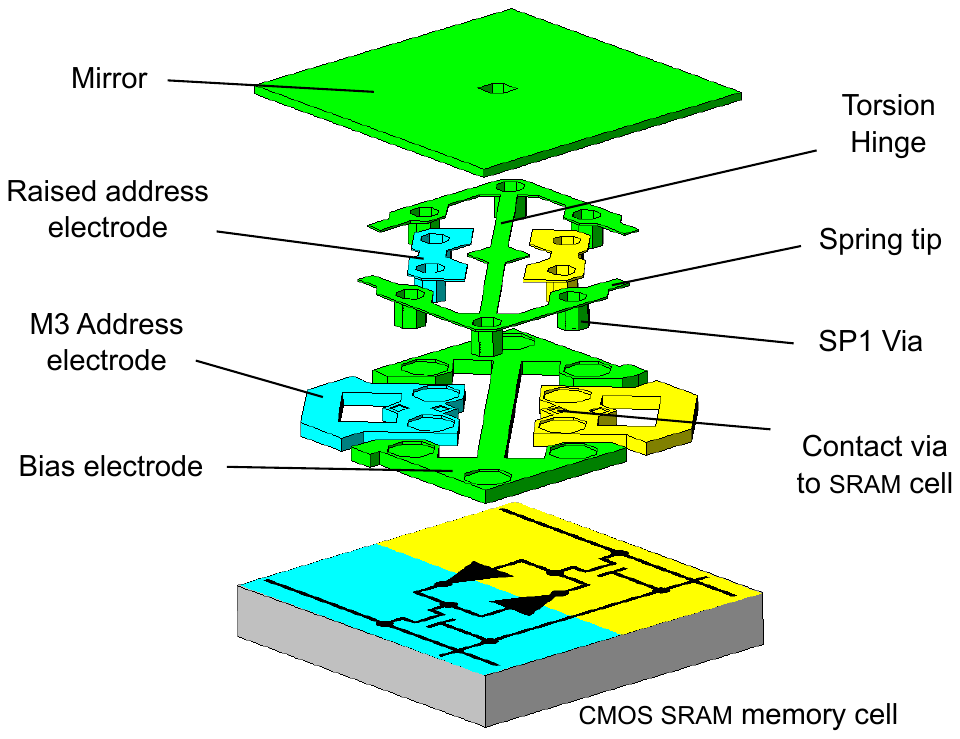}}
    \caption{Exploded diagram of the main elements of a DMD micromirror.\cite{gong_cmos_2014}}
    \label{fig:DLP}
\end{figure}

DMDs from Texas Instruments represent an extremely successful application of MEMS for digital light processing projection displays. A DMD is a planar array of aluminum mirrors that can be individually tilted in two opposite directions. The mirrors and their supporting mechanical structures are directly built on a CMOS memory circuit, each mirror being served by a dedicated pair of memory elements (Figure \ref{fig:DLP}) in complementary states (either logic 1 and 0, or vice versa). The polarity determines the tilt direction that the mirror will take when a clocking ``reset'' pulse is applied. Once the mirror has landed, the cell polarity can be maintained or reversed without affecting the tilt angle until a new clocking ``reset'' pulse is applied. 

TI commercializes several families of devices (``chipsets'') in various format. For example, DMDs with the standard 1080p format ($1920\times1080$ pixels) are offered at present in 13 different versions with 5.4$\mu$m, $7.6~\mu$m, or $10.8~\mu$m mirror pitch. DMDs with the larger DCI (Digital Cinema Initiatives) format have $4096 \times 2176$ mirrors with 5.4$\mu$m pitch. 
Another line of chipsets is represented by the DLP Cinema\textsuperscript{\textregistered} DMDs. The 1.25 Cinema 2K model with $2048\times1080$ format has the largest mirror size, 13.7$\mu$m pitch, and represents at present the best choice for astronomical applications. The double format Cinema 4K with $4096\times 2160$ provides higher resolution but at the price of a smaller 7.6$\mu$m mirror size. Cinema devices are specifically designed for the digital movie industry and generally only available to selected partners and customers under special agreements. Other groups have investigated different approaches to the design of programmable slit masks working in reflection \cite{noauthor_sandia_nodate, zamkotsian_moems_2017} but none has yet reached a level of maturity beyond early prototype. 



\subsection{Performance comparison}
\subsubsection{Mechanical and electrical reliability} 
The transmissive nature of the MSAs requires the microshutter blades to fully open with a 90\textdegree\, rotation angle to allow light from a selected target to reach the spectrometer.
The large rotating angle and the capability of achieving fully open and close settings during operations pose significant engineering challenges, aggravated by the 30~K operating temperature requirement of JWST/NIRSpec.  The mechanical design of the shutter blades was the focus of the early development efforts at University of Maryland \cite{fettig_aspects_1999} and results were highly encouraging. Still, the MSAs selected for flight on JWST/NIRSPEC present a significant fraction of non-functional shutters. During the AIV (Assembly, Integration and Verification) phase of the MSAs subsystem, partially open shutters may be covered with aluminum plugs prior to their final integration to prevent them to become a source of parasitic light\cite{chambers_cryogenic_nodate}, but 
shutters failing at a later stage will remain defective and those stuck in open state represent a source of light leakage. The fraction of functional shutters of JWST/NIRSPEC soon after launch has been estimated to be 82.5\% \cite{rawle_-flight_2022}, with a decrease of 1.2\% in the following 20 months \cite{bechtold_nirspec_nodate}. Instability is also an issue, as the JWST/NIRSpec maps show a population of shutters that malfunction randomly, complicating the planning and execution of successful observations.
Another type of malfunction impacting observing efficiency is represented by the occasional electrical shorts producing glow strong enough to make exposures suboptimal for scientific investigation \cite{bechtold_nirspec_nodate}. The NGMSA development aims at addressing these issues.

DMDs, on the other hand, have torsion bars tilt in the range 10\textdegree-17\textdegree. Since projection applications require mirrors to be modulated at several kHz between the ON/OFF states, achieving high reliability was a major goal in the early DMD development phases but already in the early years 2000's this was considered a solved issue \cite{douglass_dmd_2003}. In general, DMDs perform flawlessly and failed mirrors are a rare occurrence in commercial devices. Recent analysis indicates that even at their resonant frequency (in the range of hundreds of kilohertz), commercial DMDs show consistent performance surpassing $2\times 10^{10}$ oscillation cycles\cite{tong_continuous_2025}.

\subsubsection{Environmental Conditions}
MSA are flying currently on JWST/NIRSpec and can be  therefore considered at NASA TRL-9. A 128x64 NGMSA array prototype has flown on the FORTIS sounding rocket mission launched in Oct. 2019 \cite{li_successful_2020} and reaching an advanced TRL level. 
DMDs have been extensively tested on the ground. The first test campaigns were performed in Europe when DMDs were briefly considered by ESA as candidates for what later became the Euclid mission \cite{cimatti_space_2009, robberto_overview_2008}. These tests included operability at low temperatures, total ionizing dose, thermal cycling and vibrations and confirmed the DMDs viability for space applications \cite{zamkotsian_space_2010,
zamkotsian_successful_2011}. More recently, NASA has funded an extensive campaign aimed at space qualifying commercially available DMDs with 13.7~$\mu$m pixel pitch\cite{Travinsky2018EvaluatingSpectrometers}.  These campaigns included
proton testing \cite{Fourspring2013}, heavy-ion radiation testing\cite{TravinskyPerformance}, gamma ray testing\cite{oram_effects_2019}, 
shock and vibrations\cite{VorobievManufacturi} and low temperature testing\cite{Travinsky2018EvaluatingSpectrometers}. Waiting for a flight demonstration, TI DMDs can be considered at TRL-5\cite{vorobiev_spectroscopic_2021}.

\subsubsection{Slit size}
The physical size of a single MSA shutter, 100$\times200~\mu$m center-to-center, is well matched to  8~m class telescopes like JWST (diameter 6.5m) operating at the diffraction limit in the near-IR. The f/16.67 native focal ratio of the JWST Optical Telescope Assembly, reduced to about f/12.5 by the NIRSPEC optics, results in shutters with $0.20''\times0.46''$ open area, nicely tailored to the telescope diffraction limit in the near-IR ($\lambda/D=0.1''$ at 3.15~$\mu$m, about the central wavelength of the NIRSPEC spectral range).

\begin{figure}[tph!]
\centerline{\includegraphics[totalheight=7.5cm]{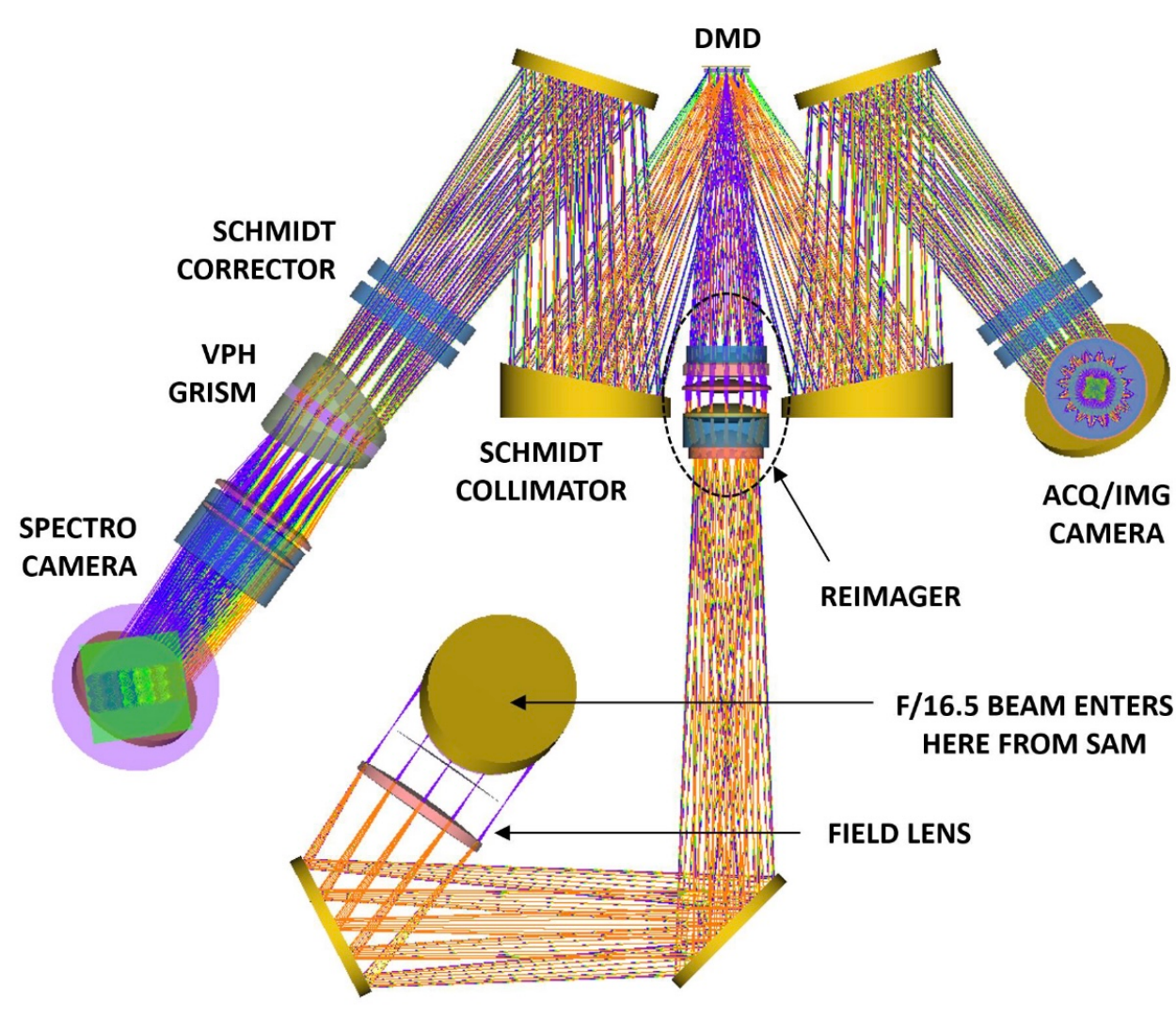}}
    \caption{Top View of the SAMOS optical layout, showing how the diameter of the reimaging optics, and therefore the f/\# of the beam illuminating the DMD, is constrained by the mirror tilt angle\cite{smee_opto-mechanical_2018}. }
    \label{fig:SAMOS}
\end{figure}

The much smaller format of DMD mirrors requires faster optics to achieve comparable scale on the sky with 8~m class telescopes. In the case of JWST, projecting 0.2'' on a 0.137$\mu$m aperture would require illuminating the DMD at f/2.1. There is, however, a limit to the speed of the beam illuminating the DMD chip (along its normal axis) set by the mirror tilt angle, as one needs to maintain the separation between incoming and reflected beams. For a $24^\circ$ angle this limit is in theory f/2.35, but in reality it becomes rapidly impractical to significantly exceed the f/4 beam adopted by SAMOS\ref{fig:SAMOS}. This value would correspond to a scale 0.108'' per micromirror on a 6.5m aperture telescope, adequate for diffraction limited performance at visible wavelengths. Moving to longer wavelengths, a PSF overfilling the slit results in diffraction effects that degrade performance. 
Designs have been proposed that illuminate the DMD from the diagonal to increase the beam separation and therefore use faster f/\# \cite{content_offspring_2008,spano_dmd_2009,content_atlas_2018,content_sirmos_2024}, but at this time no astronomical instrument has been built with DMD optics faster than f/4 (IRMOS f/4.7\cite{winsor_optical_nodate}, RITMOS f/7.6 \cite{meyer_ritmos_2004}). We underline that in the early development times of JWST/NIRSPEC, NASA started a program to develop DMDs similar to the early TI devices with 100$\mu$m as mirror size, the same value also eventually adopted by the MSA. 
A new generation of MMDs should have larger mirror size, 100~$\mu$m being currently the target for the HWO program. 

The tilt angle also represents a parameter that can be optimized.
Larger tilt angles allow for bigger separation of the incoming and reflect beams and therefore faster optics. At present TI commercializes DMD with tile angles as large as $17^\circ$, for the DLP3010 with 5.4~$\mu$m mirror size.

\subsubsection{Number of shutters}
The number of shutters comprised in a single MEMS unit, either MSA or DMD, is a key performance parameter driving the number of targets that can be simultaneously observed and, together with the slit size, the areal coverage of the spectrograph. 

The JWST/NIRSpec MSAs have $365\times 171$ format corresponding to 62,415 shutters. This is a relatively modest value, but due the large size of the individual shutters a single NIRSpec MSA covers a sizeable projected area on sky, about $70''\times70''$. Moreover, MSA can be butted together accepting gaps that, in the case of NIRSpec, are about $17''$ and $37"$ in the dispersion and cross-dispersion directions, respectively. As a result, the $2\times2$ MSA mosaic of JWST/NIRSpec extends over a $3.6'\times 3.4'$ area on the sky.

By comparison, the 2048 x 1080 format of the TI CINEMA 2k devices corresponds to more than 2.2 million mirrors, i.e. 35.4 times more elements than a single MSA unit. Illuminated at f/4 by a 6.5~m JWST-like primary mirror, a single Cinema DMD covers a field about $3.7'\times2.0'$. The packaging of TI DMDs, however, prevents butting together DMDs unless one accepts very large gaps. A $2\times2$ mosaic would result in a layout with very low fill-factor fed by large optics, an undesirable solution. A new generation of DMDs with larger mirrors should include the possibility of butting together multiple units, following the same strategy adopted for the MSAs.

\subsubsection{Fill factor and throughput}
 
Together with the number of shutters, it is important to consider the fill factor of individual shutters, another key factor affecting observing efficiency. In the case of MSA,  the $105~\mu$m$\times 204~\mu$m center-to-center periodicity of the support structure results in $78~\mu$m$\times178~\mu$m open areas, due to the frame thickness and the need to install light shields to prevent light leaking through the blade's edges. As a result, the geometric fill factor is about 65\% \cite{jakobsen_near-infrared_2022}.

The Cinema 2K DMDs, on the other hand, have $13.7\mu$m pitch center-to-center vs. a $13.0~\mu$m mirror size, resulting in a 88\% geometric fill factor for perpendicular illumination and taking into account the 12$^\circ$ mirror tilt. Larger values, up to 97\% are reported by TI for other models with smaller mirror size and larger tilt angles. 
One can see how the different strategies, transmission or reflection, drive the fill factor: for transmissive devices the need of having a mechanically robust support structure and optimal rejection of stray light drives down the fill factor. For reflective devices, the substrate does not interfere with the light path and rejection of undesired photons drives the design to the smallest possible gaps between mirrors. Both factors push the DMD fill factor to high values.

Regarding throughput, as already mentioned the MSAs have 100\% transmission. TI DMDs instead utilize a proprietary multilayer Al coating aimed at ensuring long life-time of the mirrors oscillating at high speeds. Coating the surface of the DMD mirrors e.g. to increase throughput at UV wavelengths has been successfully demonstrated \cite{Travinsky2018EvaluatingSpectrometers}.
Moreover, DMDs are protected by a window with AR coating optimized for transmission in the 420~nm–680~nm wavelength range, with 96\% estimated transmission in the double pass. TI commercializes a few DMD models with window coating optimized for either visible, near-ultraviolet or near-infrared wavelengths. While custom window replacement of TI DMDs has also been successfully demonstrated \cite{TravinskyPerformance}, a future generation of DMDs must provide the flexibility of optimizing both mirror and window coatings during construction in order to maximize throughput over the spectral range of interest without risks of damaging the devices.

\subsubsection{Contrast}
In the early JWST days major attention was paid to the issue of contrast, intended as the ability of a MEMS-based slit mask mechanism to reject light from unwanted point sources. Calculations focused on the astrophysical problem of determining the density of sources bright enough to degrade, through their leaking light, the requirement on the signal-to-noise for nearby faint targets. The selection of MSAs over the DMDs was largely based on the expectation that MSA could provide superior contrast, surpassing more easily the 2000:1 requirements set by astronomers. 
Through the years the issue of contrast of DMDs has been the focus of several studies with a broad variety of values reported by various teams. Planning to return more in detail to this point in a future paper, we underline here just a few basic considerations.

Contrast depends on the type of illumination. Determining the fraction of photons leaking through the MEMS from a bright point source captures just one possible source of contamination. This light will create a ghost image of the unwanted bright target, and the corresponding ghost spectrum if a dispersing element is introduced in the light path. The brightness of the ghost image normally depends on the f/\# of the illumination, that therefore affects also the light scattered and/or diffracted around. Whereas the light scattered off-beam by as single source can be negligible, when uniform illumination (e.g. sky background) is present every point of the field becomes a significant source of stray light. A point-like source spread over multiple slits represents an intermediate case, as the diffraction between the various elements receiving coherent light has to be accounted for \cite{Piotrowski2022OpticalDevice, Piotrowski2023OpticalDevice}. Other factors to consider are the leaks from defective, especially fully open, slits/shutters that in the case of a spectrograph produce spectra of the exposed background contaminating entire rows. From this point of view, a device  ``point source'' contrast rejection has to be evaluated together with the  ``cosmetic quality'' of the device. 

For the MSAs, an average contrast 66,000:1 was originally reported \cite{kutyrev_microshutter_2008}, but for the same MSA devices flying on the JWST/NIRSPec MSA the cumulative MSA flux leakage is actually measured at a level close to one part in ~$\sim50$ (2\%)\cite{NIRSpecMSA}. 
For DMDs with the same 13.7~$\mu$m mirror size of the the 1.2 Cinema DMD, measures performed at NASA/GSFC resulted in values between 6000:1 and 25,000:1 depending on the experimental setup\cite{navarro_measurements_2016}. 
Measures of the contrast on-sky with SAMOS are still subject of investigation as the faintness of the sky background makes the leaking flux nearly indistinguishable from the noise floor level. 

\section{Development of new Micro-Mirror Devices}
The overarching goal of the proposed research is to develop a micromirror array that is mechanically robust and optimized for the MOS that will enable unprecedented capabilities for advancing future astronomic exploration.  Our target micro mirror device is expected to meet the specifications listed in Table 1.

\begin{table}[ht]
\caption{Specifications of the targeted micro mirror device.} 
\label{tab:fonts}
\begin{center}       
\begin{tabular}{|l|l|} 
\hline
\rule[-1ex]{0pt}{3.5ex}  Mirror Size & 30 $\mu$m x 30 $\mu$m  \\
\hline
\rule[-1ex]{0pt}{3.5ex}  Tilt angle & 15\textdegree   \\
\hline
\rule[-1ex]{0pt}{3.5ex}  Array size & 1000 x 1000   \\
\hline
\rule[-1ex]{0pt}{3.5ex}  FOV at f/4 & 6.3 x 6.3 arcmin   \\
\hline
\rule[-1ex]{0pt}{3.5ex}  Wavelength & Visible to 2.5 $\mu$m   \\
\hline
\rule[-1ex]{0pt}{3.5ex}  Mirror reflectivity & 96\%  \\
\hline
\rule[-1ex]{0pt}{3.5ex}  Contrast ratio & 1:10,000 (blue), 1:5000 (red)  \\
\hline
\rule[-1ex]{0pt}{3.5ex}  Reconfiguration time & \textless 4 seconds  \\
\hline
\end{tabular}
\end{center}
\end{table} 


The Micro-Mirror device (MMD) will consist of an addressable array of micromirrors at a 30 $\mu$m pitch.  The array size is configurable for the necessary telescope application to cover the optical field.  Our initial target micro mirror device will consist of a micromirror array of 1000 columns and 1000 rows.  The targeted individual micromirror will be 30$\mu$m x 30$\mu$m in size with a 15° tilt angle.  The mirror will be electrostatically driven.  The mechanical robustness of the mirror will be high since the mirror only tilts 15° in comparison with microshutters that have to rotate close to 90°.  We will leverage our prior MEM design and fabrication experience to choose the right hinge and mirror material and processing parameters to ensure its long-term mirror operation.

\subsection{Pixel Operation} 

\begin{figure}[tph!]
\centerline{\includegraphics[trim={3cm 21.2cm 3cm 2.6cm}, totalheight=4.2cm]{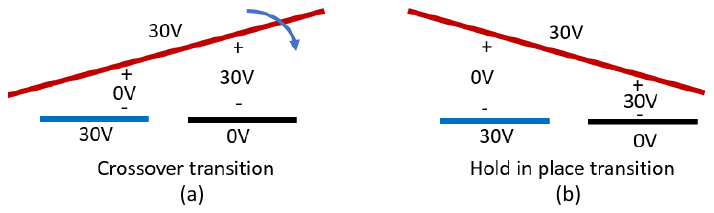}}
    \caption{Pixel operation (a) crossover transition from ON to OFF;  (b) hold in place transition staying at OFF position.}
    \label{fig:pixel op}
\end{figure}

The pixel operation in the array goes through either a crossover transition (from ON state to OFF state or from OFF state to ON state) or a hold-in-place transition.  The ON and OFF states of a pixel refer to the landed directions of the pixel with reference to the flat position of the pixel with ON state corresponding to a positive landing angle. For a pixel that will go through a crossover transition as shown in Figure \ref{fig:pixel op}a, the voltage of mirror will be set at +30V, the voltage of the electrode that is in close proximity with the mirror will be set at +30V.  The voltage of the other electrode will be set at 0V. Note the voltage values referred here are just for illustration purpose.  The electrical potential between the mirror and the mirror electrodes will be 0V and +30V respectively.  The micromirror will cross over to the side that has higher electrostatic attraction.  For a pixel that will go through a hold-in-place transition, the electrode voltages will be set opposite to those for the crossover pixel, as shown in Figure \ref{fig:pixel op}b.  The electrical potential between the mirror and the electrode in close proximity is +30V that keeps the mirror in the original landed position.

\subsection{Array sequencing control architecture} 
\begin{figure}[tph!]
\centerline{\includegraphics[trim={3cm 16.2cm 3cm 2.5cm}, totalheight=9.2cm]{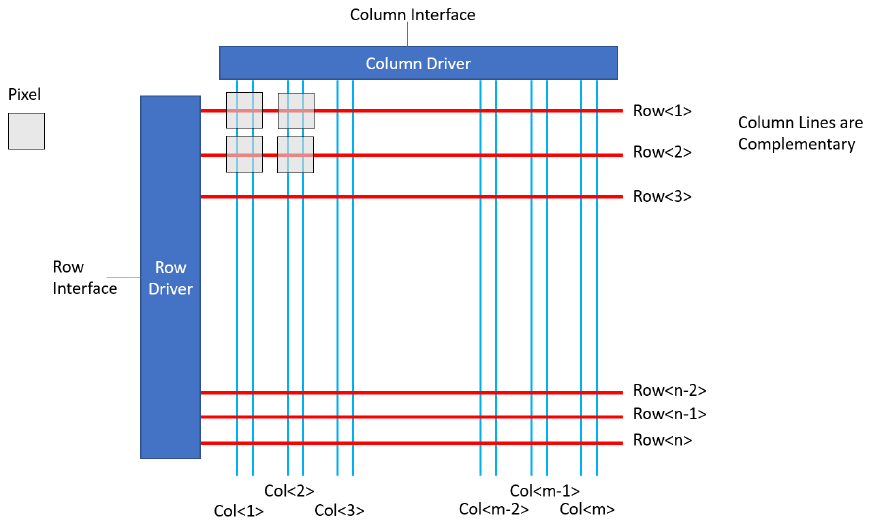}}
    \caption{Array sequencing control architecture with row and column drivers.}
    \label{fig:pixel seq}
\end{figure}

To enable individual states in the micromirror array, each micromirror row will be connected to a row select driver. The state information for each pixel will be controlled by a pair of complementary column lines that are connected to a column driver.  For this device, there will be 1000 row lines and 1000 pairs of complimentary column lines as shown in Figure \ref{fig:pixel seq}. The mirror array sequencing is accomplished on a row by row basis.  The pixel state data will be loaded serially into the column segment drivers.  The row selection will be controlled by the row segment drivers.  The sequencer will synchronize the row select and the column data. After the array has completed all transitions, the state of the pixel array will be held by the sequencer. 

\subsection{Fabrication processes and packaging} 
\begin{figure}[tph!]
\centerline{\includegraphics[trim={3cm 18cm 5cm 2.5cm}, totalheight=8cm]{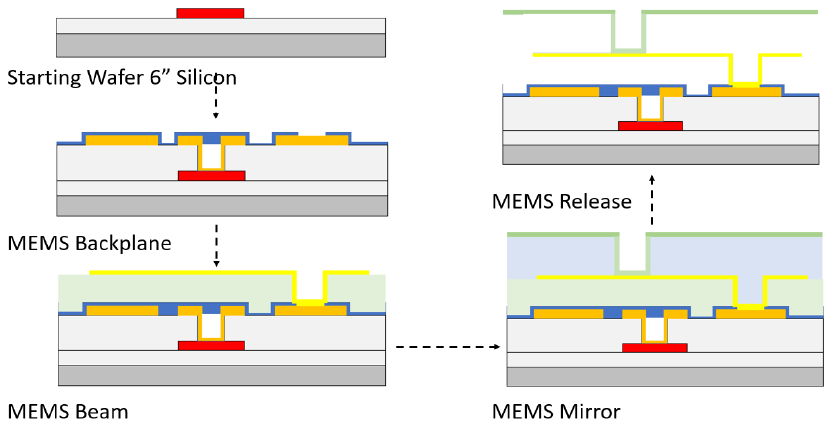}}
    \caption{MMD fabrication processes.}
    \label{fig:MMD fab}
\end{figure}

The proposed MMD will have three metal layers including Metal3, middle layer and the top mirror layer shown in Figure \ref{fig:MMD fab}.  Standard semiconductor processes will be used to fabricate the device. The process flow is provided in the following.
1)	The Metal3 layer is built directly on top of the substrate. The address electrodes and the routings to conduct the bias voltage are built in this layer.  
2)	Create first sacrificial layer, “spacer-1” using photoresist. The vias are opened up in the patterning process for spacer-1 to allow the subsequently deposited hinge metal a free path to contact the electrodes in Metal3.
3)	Deposit torsion hinge metal film.  Pattern and etch the hinge metal to create the torsion hinge, raised address electrodes, and mirror tilt “stops.”
4)	Create second sacrificial layer “spacer-2” using photoresist.  Mirror vias are opened up during the patterning process to form the connection between the mirror and the middle layer.
5)	Deposit mirror metal film.  Pattern and etch mirror metal to singulate each pixel.
6)	Sacrificial layers are removed through photoresist plasma ashing. 

The entire silicon chip will then be mounted on a rigid, thermally conductive substrate to allow for mounting and aligning in the overall system. The electronics that are responsible to configure the pattern will be mounted on the outside of the array in a row select and column data segment configuration that is similar to a Passive Matrix LCD.  The connector to the system electronics will be mounted near the row and column segment drivers to allow for proper control of the pattern on the micromirror array.  The array will be surrounded by a sized seal ring to allow for the window to be mounted over the micromirror array.  

\subsection{Testing and verification}
The testing and verification of the micro mirror devices will be conducted in 3 configurations. The first test configuration will be the wafer-level characterization in which the micromirror parameters and the overall yield of the array will be obtained.  This characterization will occur on a probe station using standard semiconductor test equipment on a camera-based system.  The wafer level characterization data will be analyzed to provide feedback for each design iteration.  
The second test configuration is the package level characterization, which will occur after each package learning cycle has completed.  Package level characterization can be completed on the probe station using the packaged devices or in a socket using the completed packaged devices.  This test will allow for environmental and reliability testing, as well as micromirror characterization and yield.  This characterization can be performed on a probe station using standard semiconductor test equipment, a heated chuck (test fixture) and a camera system.  
The third test configuration is the system level characterization that will be conducted in the actual system with the system level electronics. This test will allow for system characterization and verification. System level characterization and verification data will be analyzed to verify system specification compliance for the micro mirror devices. 

\section{Summary and Conclusion}
We introduce our new program to develop "Micro-Mirror Devices" (MMD), two-dimensional MEMS arrays of individually addressable micro-mirrors specifically optimized for astronomy. We are motivated by both the success history of MEMS devices as slit selection mechanism for multi-object spectroscopy and the limitations of the currently available devices, that we have elucidated in Sec. 2. In Sec. 3 we have presented our program aimed at delivering sequential generations of devices, starting with our first baseline devices with 30\textmu m x 30\textmu m pixel size arranged in a 1K x 1K format with tilt angle 15 degrees. Guided by the current requirements for the HWO project, our goal is to bring to maturity a technology capable of delivering arrays of 2K x 2K element of 100\textmu m x 100\textmu m, buttable on two sides to achieve even larger formats reaching TRL-5 by 2029.  


\bibliography{report} 
\bibliographystyle{spiebib} 

\end{document}